\documentclass[12pt,onecolumn,draftclsnofoot]{IEEEtran}
\IEEEoverridecommandlockouts
\usepackage{color}

\usepackage{times}
\usepackage[final]{graphicx}
\usepackage{amsmath,amsfonts}
\usepackage{amsthm}
\usepackage{amssymb,amsbsy}

\usepackage{cite}
\usepackage{multirow}

\newcommand{\ignore}[1]{}

\newcommand{\hsp}{\hspace{0.1in} }
\newcommand{\hspp}{\hspace{0.05in} }

\newsavebox{\savepar}

\begin{document}
\title{Millimeter Wave MIMO Prototype: Measurements and Experimental Results}
\author{\large Vasanthan Raghavan, Andrzej Partyka, Ashwin Sampath,
Sundar Subramanian, Ozge Hizir Koymen, Kobi Ravid, Juergen Cezanne,
Kiran Mukkavilli, Junyi Li
\thanks{The authors are with Qualcomm Corporate R\&D, Bridgewater, NJ 08807, USA
and Qualcomm Corporate R\&D, San Diego, CA 92121, USA.}}

\maketitle
\vspace{-10mm}

\begin{abstract}
\noindent
Millimeter-wave multi-input multi-output (mm-Wave MIMO) systems are one of the candidate
schemes for 5G wireless standardization efforts. In this context, the main contributions
of this article are three-fold. 1) We describe parallel sets of measurements at identical
transmit-receive location pairs with $2.9$, $29$ and $61$ GHz carrier frequencies in
indoor office, shopping mall, and outdoor settings. These measurements provide insights
on propagation, blockage and material penetration losses, and the key elements
necessary in system design to make mm-Wave systems viable in practice. 2) One of these
elements is hybrid beamforming necessary for better link margins by reaping the
array gain with large antenna dimensions. From the class of fully-flexible hybrid
beamformers, we describe a robust class of directional beamformers towards meeting
the high data-rate requirements of mm-Wave systems. 3) Leveraging these design
insights, we then describe an experimental prototype system at $28$ GHz that realizes
high data-rates on both the downlink and uplink and robustly maintains these rates in
outdoor and indoor mobility scenarios.
In addition to maintaining large signal constellation sizes in spite of radio
frequency challenges, this prototype leverages the directional nature of the
mm-Wave channel to perform seamless beam switching and handover across mm-Wave
base-stations thereby overcoming the path losses in non-line-of-sight links
and blockages encountered at mm-Wave frequencies.
\end{abstract}


\section{Introduction}
\label{sec1}
Millimeter-wave multi-input multi-output (mm-Wave MIMO) systems are one of the
candidates for the physical layer (PHY) in the currently ongoing standardization efforts for
the Fifth Generation ($5$G) air link specifications. Over the last few years, there has
been an exploding interest on mm-Wave systems (see,
e.g.,~\cite{boccardi1,rappaport,heath_sp_overview} and references therein). Most of
these works either present expectations from mm-Wave systems,
or use-case analysis, or channel measurements, or performance studies assuming certain
PHY abstractions. With this backdrop, the focus of this paper is on understanding the
implementation of mm-Wave systems in practice and to present a complete picture
starting from channel measurements to system design implications to prototype performance.

Towards this goal, we first study electromagnetic propagation in
the mm-Wave regime using a number of parallel measurements at the {\em same} transmit-receive
location pairs in different use-cases (indoor office, shopping mall and outdoor) at $2.9$,
$29$ and $61$ GHz. Such a parallel set of measurements minimizes the number of
confounding factors and allows a direct comparison of propagation across different
carrier frequencies. A limited number of such measurement studies at the same
location pairs are available in the literature.
While our studies show that losses at mm-Wave frequencies
are typically higher than with sub-$6$ GHz systems, these losses are not substantially
worse. Nevertheless, additional losses due to hand/human blockages and material
penetration can be significantly detrimental to the link margins and are expected to
play the role of a serious differentiator for a mm-Wave chipset solution.

The above observations motivate the use of beamforming to overcome these losses.
We then briefly describe the radio frequency (RF) component challenges that impact
the design of a practical mm-Wave chipset. The use of near-optimal beamforming structures
needed to improve the link margin in both single- and multi-user contexts in a
practical mm-Wave chipset is not viable due to cost and complexity challenges of radio
frequency (RF) components. Thus, secondly, to overcome these constraints, we are
motivated to use a certain subset of {\em directional} beamformers
for MIMO transmissions. In addition to low complexity, the proposed
approaches also enjoy advantages such as robustness to phase changes
across paths (an issue of immense importance for small wavelength systems such as mm-Wave)
and a simpler system design for initial user equipment (UE) discovery and subsequent
beam refinement.

With this background, thirdly, we describe our prototype system operating at $28$ GHz
that realizes a robust directional beamforming solution leading to high data-rates on
both the downlink and uplink. We describe various experiments performed with the
prototype in both outdoor and indoor scenarios
and provide unique insights into the operations of a practical mm-Wave system.
The key elements tested by these experiments include robustness of mm-Wave links in
non-line-of-sight (NLOS) settings via beam switching in response to mobility and
blockage, inter-base-station handover, and interference management. Comparable
prototypes in the literature such as~\cite{roh} mostly emphasize peak throughputs
in line-of-sight (LOS) settings and do not provide lessons applicable for practical
deployments. Other prototypes of importance in practical
deployments include the CAP-MIMO architecture~\cite{brady_tcom}
and~\cite{rui_zhang,lens_array1,lens_array2} that apply lens array techniques
for steering multiple beams from the base-station.

\section{Millimeter-Wave Channel Measurements and System Implications} 
\label{sec2}
The focus of this section is on reporting mm-Wave channel measurements at $2.9$, $29$
and $61$ GHz, which are representative of the three most-likely commercial offerings
in the $2018$-$20$ time-frame and likely to be compared against each other in terms of
performance: sub-$6$ GHz 5G-NR, mm-Wave MIMO 5G-NR, and 802.11ad/ax/ay. While a number
of mm-Wave
channel measurement campaigns have been reported in the literature, the novelty of this
work is on channel propagation comparisons across these three carrier frequencies at
{\em identical} transmit-receive location pairs in different use-cases.
Such studies are important as they eliminate most confounding factors that prevent a
direct comparison across frequencies.

Towards this goal, channel sounding is performed with both omni-directional antennas as well as
directional horn antennas. For directional measurements, an azimuthal scan (or
a $360^{\sf o}$ view) with a $10$ dBi gain horn antenna and producing $39$
directional slices, and a spherical scan ($360^{\sf o}$ azimuth view and
$-30^{\sf o}$ to $90^{\sf o}$ view in elevation) with a $20$ dBi gain horn
antenna and producing $331$ directional slices are generated. The time-resolution
of the channel sounder is approximately $5$ ns. An Agilent E8267D signal generator
is used to generate a pseudo-noise (PN) sequence at a chip rate of $100$ Mc/s, which
is then used to sound the channel. At the receiver, an Agilent N9030A signal
analyzer is used for acquisition and the PN chip sequence is despread using a
sampler at $200$ MHz and with $16$ bit resolution.

These sounding measurements are obtained for the indoor office setting (two floors of the Qualcomm
building in Bridgewater, NJ), indoor shopping mall setting (Bridgewater Commons Mall,
Bridgewater, NJ), outdoor settings (open areas
outside the Qualcomm building), including the suburban setting (residential location
in Bedminster, NJ) and the Urban Micro setting (New Brunswick, NJ), etc., as well as
emulation of stadium deployments. These measurement scenarios are representative of
typical applications considered for future deployment efforts.

While macroscopic channel properties such as path loss are studied with omni-directional
scans, other properties such as delay spread, path diversity, etc., are studied with
both omni-directional and directional scans. Processing of these measurements lead to the
following observations and implications on system design for mm-Wave channels. More
technical details on these studies can be found in~\cite{vasanth_tap2018}.

\noindent {\bf \underline{Path Loss:}} Measurements in different deployments are
used to study macroscopic properties of LOS and NLOS links. We use a frequency-dependent
path loss model with a close-in free space reference distance of $d_0 = 1$ m where the
path loss (in dB) at a distance of $d$ m is modeled as
\begin{eqnarray}
{\sf PL}(d) = {\sf PL}(d_0) + \alpha \cdot 10 \log_{10}(d/d_0) + {\sf X},
\hsp \hsp  {\sf X} \sim {\cal N}(0, \sigma_{\sf X}^2).
\label{eq_pathloss}
\end{eqnarray}
The path loss exponents (PLEs), denoted as $\alpha$, and the shadowing factors
($\sigma_{\sf X}$) for different types of links (LOS/NLOS) in different use-cases
are learned with a least-squares fitting of the model in~(\ref{eq_pathloss}) to the
measured data. These parameters are listed in Table~\ref{table_pathloss} and they
show that PLEs and shadowing factors for NLOS links generally increase with frequency.
For LOS links, PLEs are generally smaller than those for NLOS links and in indoor
settings can be smaller than the freespace PLE of $2$. A plausible explanation for this
observation is {\em waveguide effect} where long enclosures such as walkways/corridors,
dropped/false ceilings, etc., tend to propagate electromagnetic energy via alternate
modes/more reflective paths decreasing the PLE. Shadowing factors show inconsistent
behavior with frequency. From Table~\ref{table_pathloss}, we conclude that while
mm-Wave systems experience higher path losses than sub-$6$ GHz systems, the
differential impact of the PLEs and shadowing
factors on link margin at higher carrier frequencies is {\em \underline{not}} dramatic.

\begin{table*}[htb!]
\caption{Path Loss Model Parameters in Different Use-cases}
\label{table_pathloss}
\begin{center}
\begin{tabular}{
|c|  |c|c|c||c|c|c|  |c|c|c||c|c|c||}
\hline
${\sf Parameter \downarrow}$
& \multicolumn{3}{c||}{ ${\sf LOS}$}
& \multicolumn{3}{c||} {${\sf NLOS}$}
& \multicolumn{3}{c||}{ ${\sf LOS}$}
& \multicolumn{3}{c||} {${\sf NLOS}$}
\\ \hline \hline
${\sf f_c \hspp (in \hspp GHz) \rightarrow}$ & $2.9$ & $29$ & $61$ & $2.9$ & $29$ & $61$
& $2.9$ & $29$ & $61$ & $2.9$ & $29$ & $61$
\\ \hline \hline
&
\multicolumn{6}{|c||}{ ${\bf Indoor \hspp office}$ }
& \multicolumn{6}{|c||}{ ${\bf Indoor \hspp shopping \hspp mall}$ }
\\ \hline
${\sf PLE} \hspp (\alpha)$ & $1.62$ & $1.46$ & $1.59$ & $3.08$ & $3.46$ & $4.17$
& $1.93$ & $1.98$ & $2.05$ & $2.61$ & $2.76$ & $2.98$
\\ \hline
$\sigma_{\sf X} \hspp {\sf (in \hspp dB)}$ &
$5.49$ & $4.25$ & $4.81$ & $6.60$ & $8.31$ & $13.83$
& $5.32$ & $3.56$ & $4.29$ & $9.08$ & $9.47$ & $12.86$
\\ \hline \hline
&
\multicolumn{6}{|c||}{ ${\bf Urban \hspp Micro \hspp street \hspp canyon}$ }
& \multicolumn{6}{|c||}{ ${\bf Outdoor \hspp open \hspp areas}$ }
\\ \hline
${\sf PLE} \hspp (\alpha)$ & $2.18$ & $2.19$ & $2.22$ & $2.95$ & $3.07$ & $3.27$
& $2.41$ & $2.73$ & $2.83$ & $3.01$ & $3.39$ & $3.42$
\\ \hline
$\sigma_{\sf X} \hspp {\sf (in \hspp dB)}$ &
$4.41$ & $4.37$ & $4.84$ & $7.82$ & $8.16$ & $10.70$
& $4.60$ & $5.73$ & $6.78$ & $4.00$ & $8.03$ & $1.97$  \\ \hline \hline
\end{tabular}
\end{center}
\end{table*}

\noindent {\bf \underline{Delay Spread:}} The delay spread of the channel is an
important metric to understand the system overhead (in terms of the cyclic prefix
length for a multi-carrier design). In this context, frequency-dependent delay
spreads are observed in NLOS settings both with omni-directional and directional
antennas. While omni-directional delay spreads are small in most scenarios (for
example, the essential spread is on the order of $30$-$50$ ns in indoor office,
$50$-$90$ ns in indoor shopping mall and $150$-$300$ ns in outdoor street canyon
settings), there are also scenarios where a significantly large delay spread is
seen (e.g., even up to $800$ ns in outdoor open square settings). These extreme
scenarios can be explained with the {\em radar cross-section effect}, where
seemingly small objects that do not participate in electromagnetic propagation at
lower frequencies show up at higher frequencies. Such behavior happens as the
wavelength approaches the roughness of surfaces (e.g., walls, light poles, etc.).
Supporting these extremes without incurring a high fixed system overhead is important.
In most indoor scenarios, sparse scattering implies that the beamformed delay spread
is comparable with the omni-directional delay spread. While the same trend holds for
most scenarios in the outdoor setting, the beamformed delay spread can be
significantly smaller than the omni-directional delay spread for the tail values.

\begin{figure*}[htb!]
\begin{center}
\begin{tabular}{cc}
\includegraphics[height=1.5in,width=2.4in] {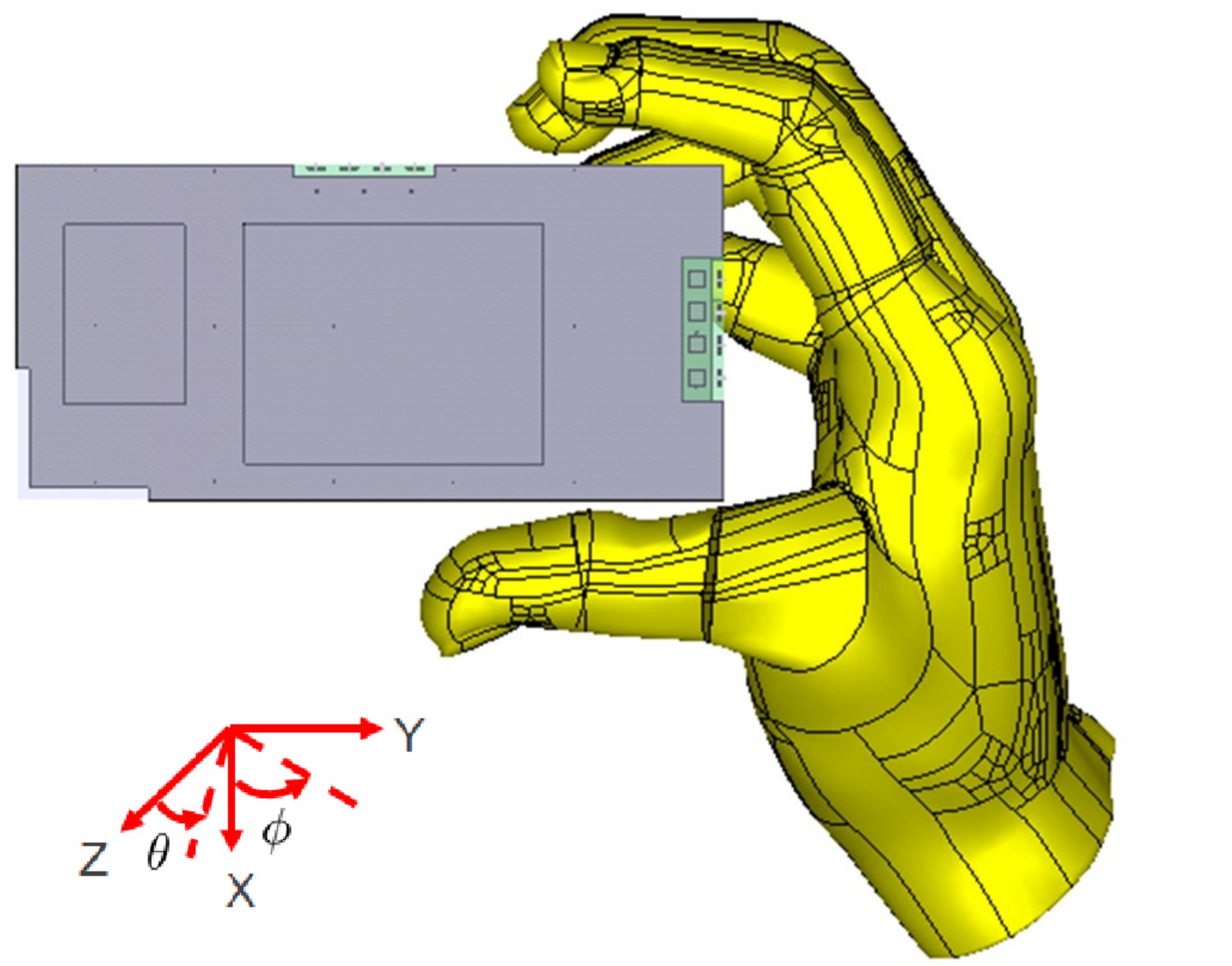}
&
\includegraphics[height=1.5in,width=3.4in] {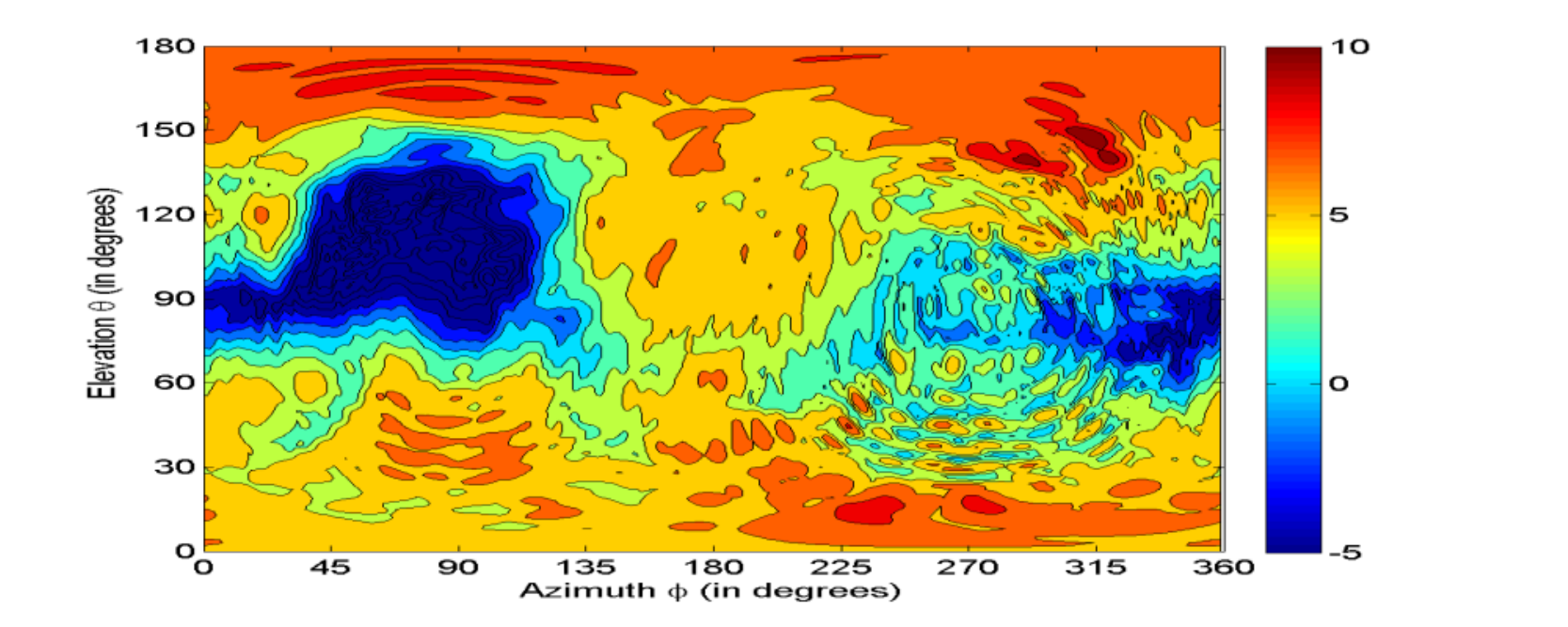}
\\
(a) & (c)
\\
\includegraphics[height=1.5in,width=3.4in] {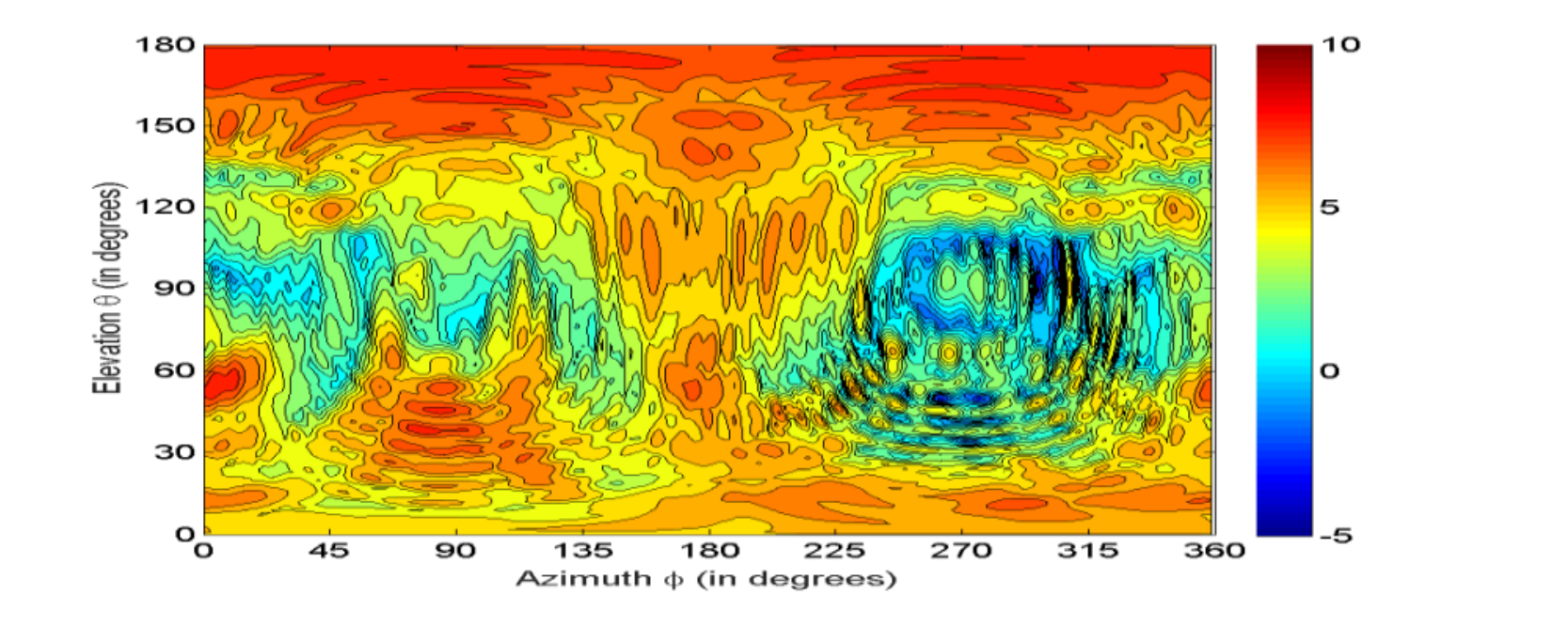}
&
\includegraphics[height=1.5in,width=3.4in] {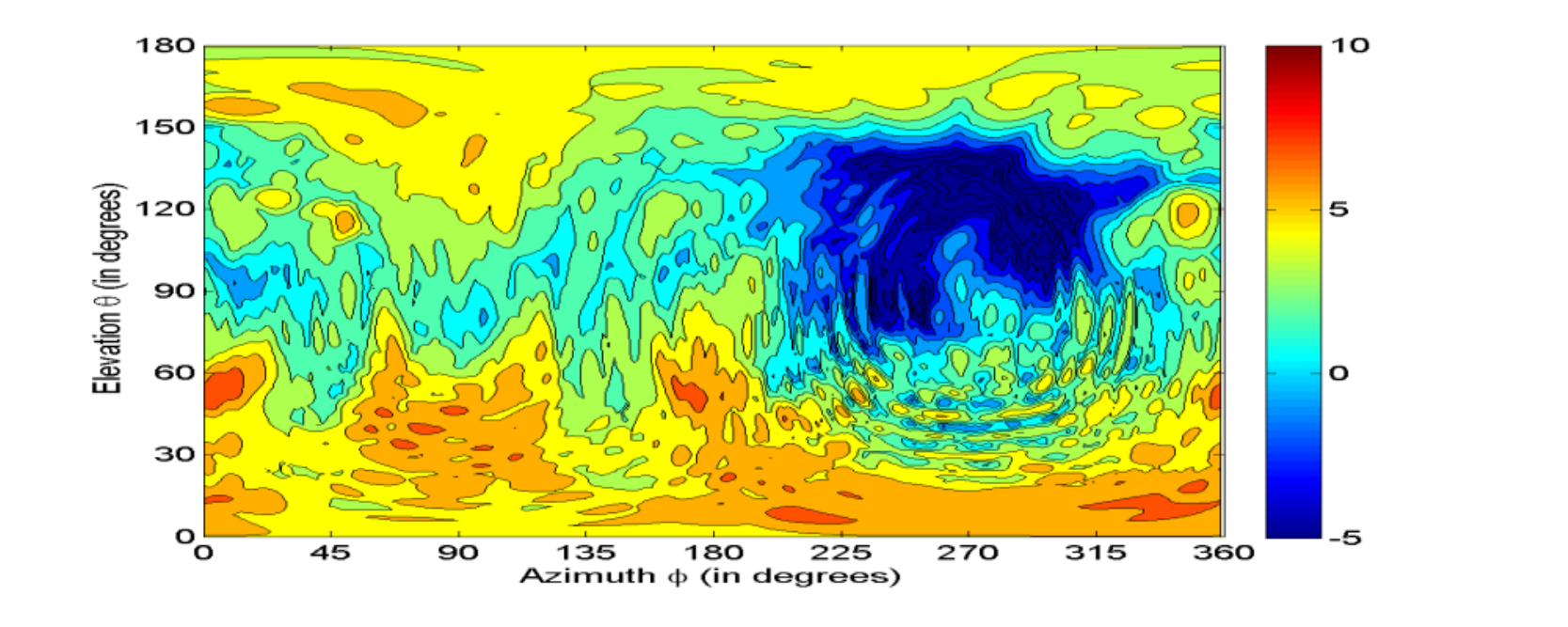}
\\
(b) & (d)
\end{tabular}
\caption{\label{fig_combined1}
(a) A typical UE design with multiple subarrays (Top and Long edges) in
Landscape mode. Received
gain as a function of azimuth and elevation angles for the UE design at $28$ GHz
in (b) Freespace mode, and with hand blocking in (c) Landscape and (d) Portrait modes.}
\end{center}
\end{figure*}

\noindent {\bf \underline{Blockage:}} An important feature that makes mm-Wave
propagation significantly different from propagation at sub-$6$ GHz frequencies
is that a large area of the UE can be easily covered and blocked by parts of
the human body, other humans, vehicles, foliage, etc. Additional link impairments
due to these blockers (not seen at sub-$6$ GHz) are observed at mm-Wave frequencies and
the practical viability of mm-Wave systems are more dependent on blockage than the
path losses reported in Table~\ref{table_pathloss}. For example, a typical UE
design with multiple linear subarray units of four antenna elements (on the Top
and Long edges of the device) is presented for the Landscape mode in
Fig.~\ref{fig_combined1}(a). Corresponding to this UE antenna design,
Figs.~\ref{fig_combined1}(b)-(d) show the received gain in azimuth and elevation
in Freespace, and with hand blocking in the Landscape and Portrait modes, respectively.
While almost the entire sphere is covered around the UE in the Freespace mode,
the presence of hand leads to an angular blockage region of $160^{\sf o} \times
75^{\sf o}$ (blue areas) in the Landscape mode. The region in blue stretches from
behind the palm to the thumb. In this setting, the Top edge subarray is not useful
and the Long edge subarray allows good signal reception. Furthermore, in the
Portrait mode, a blockage region of $120^{\sf o} \times 80^{\sf o}$ (blue areas) is
seen and the Long edge subarray does not play an important role in signal
reception as it is blocked with the fingers resulting in significantly
deteriorated antenna efficiencies. However, the Top edge subarray is not affected
much with the presence of the hand. These observations suggest that subarray diversity
in UE design is critical in overcoming near-field obstructions as well as to ensure
coverage at the UE side over the entire sphere.

\noindent {\bf \underline{Penetration Loss:}} For the outdoor-to-indoor coverage
scenario, material measurements show that penetration loss generally
increases with frequency. Further, periodic notches that are several GHz
wide and often with more than $30$ dB in loss are seen. These losses are
attributed to changing material properties with frequency due to which
signals constructively/destructively interfere from different surfaces
that make the material. While a similar trend is observed across these
experiments for both polarizations and different choices of incidence
angles, the precise loss at a frequency and the depth of the notches
depend on the material, incidence angle and polarization. This observation
motivates the need for designs that support both frequency and spatial diversity.

\noindent {\bf \underline{Path Diversity:}} Low pre-beamforming
signal-to-noise ratios (${\sf SNR}$s)
are typically the norm when mm-Wave path losses and additional blockage/penetration
losses are incorporated with typical equivalent isotropically radiated power
(EIRP) constraints. Thus, a viable system design has to overcome these huge
losses with beamforming array gains from the packing of a large number of
antennas within the same array
aperture~\cite{heath_sp_overview,hur,roh,brady_tcom,sun,oelayach,raghavan_jstsp}. In this
context, a small number of (at most $4$-$6$) well-spread out (in direction)
clusters/paths rendering multi-mode/multi-layer signaling viable are typically
observed. The viability of multiple modes suggests the use of both single-user
MIMO strategies for increasing the peak rate as well as multi-user MIMO strategies
for increasing the sum-rate~\cite{sun,vasanth_jsac2017}. Such modes also offer
robustness against blockages via intra-base-station beam switching. In addition
to the likelihood of multiple viable paths to a certain base-station, there are
also viable paths to multiple base-stations. These observations suggest the criticality
of a dense deployment of base-stations for robust mm-Wave operation and inter-base-station
handover to leverage these paths. Integrated access and backhaul operation is
highly desirable for small cell deployment, which also leads to the
need to study inter-base-station interference management issues more carefully.

\section{Experimental Prototype Description}
\label{sec3}
Motivated by the system design  intuition developed in the previous section, 
we now describe our experimental prototype system operating in a time-division duplexing
framework at $28$ GHz. In this setup, baseband analog in-phase and quadrature (IQ) signals
are routed to/from the modem to an IQ modulator/demodulator at $2.75$ GHz center frequency.
The $2.75$ GHz intermediate frequency signal is translated to $28$ GHz using a $25.25$ GHz
tunable local oscillator (with a $100$ MHz step size). The bandwidth supported is $240$
MHz at a sampling rate of $240$ Msps. ADCs with an effective number of bits (ENOB)
resolution of $8$ bits are used at both ends.
At the base-station end, the $28$ GHz signal is routed to an $16 \times 8$ element
planar array (a waveguide design) and analog beamforming is applied using tunable
four bit phase shifters and gain controllers. The prototype uses a transmit power
dynamic range of $19$ dB with a maximum EIRP of $55$ dBm. As motivated earlier,
to overcome blockage, the UE end is made of four selectable subarrays,
each a four element phased array of either dipoles or patches as in Fig.~\ref{fig_combined1}(a).

With beamforming being a central component in meeting the mm-Wave link budget, RF
component and architecture-driven challenges (e.g., cost, power, complexity, form
factor, regulatory constraints, etc.) play a principal role in determining
practically viable hybrid beamforming solutions. In this context, the {\em sparse}
and {\em directional}
channel structure suggests the use of a certain subset of {\em directional}
beamforming strategies along the dominant clusters/paths at both ends~\cite{oelayach,raghavan_jstsp,brady_tcom,vasanth_jsac2017} relative to
optimal beamforming along the dominant eigen-modes/singular vectors of the channel
matrix. Directional beamforming structures offer robustness to small perturbations
in the channel matrix and also allow a tradeoff between peak beamforming gain and
initial UE discovery latency (with minimal loss relative to the optimal schemes)
via the construction of a hierarchy of directional codebooks~\cite{raghavan_jstsp}.
The directional channel structure can be leveraged for scheduling and can also be
generalized to multi-user beamforming design with the following solutions: i) beam
steering to each UE (with complete
agnosticism of the interference caused to other users), ii) zeroforcing (where
each user's beamforming vector steers a beam null to the other users), and iii)
generalized eigenvector precoding (that performs a weighted combination of beam
steering and beam nulling). These solutions can result in substantial performance
improvement over single-user solutions. The readers are referred
to~\cite{vasanth_jsac2017} for technical details on these constructions as well as
performance studies in outdoor and indoor deployments.

Motivated by the robustness of directional beamformers, this solution is
implemented in the prototype by leveraging the beam
broadening principles described in~\cite[Sec.\ IVB]{raghavan_jstsp} to construct
static analog beam codebooks to be used at both the base-station and UE ends. The
experimental system implements mm-Wave beamforming by initially determining the best
beam direction to be used at either end. After this, the system continuously evaluates
all possible beam directions from all available transmitters
and switches to the best beam and the best transmitter
(handover) with little to no performance degradation. In addition, the system adjusts
its parameters to optimize for the link type (LOS/NLOS) by ${\sf SNR}$ control allowing
upto $64$-QAM operation.

The seamless beam switching and capability to maintain high ${\sf SNR}$ enable the
experimental system to realize high rates (on both the downlink and uplink) and
robustly maintain these rates despite channel variations. That said, the main focus
behind the experimental system is {\underline{\em not}} the optimization of
data-rates, but to study the various fundamental difficulties in realizing mm-Wave
systems in practice, especially with NLOS links. In the next two sections, we
describe some experimental
results illustrating the versatility of the prototype.

\begin{figure*}[htb!]
\begin{center}
\begin{tabular}{cc}
\includegraphics[height=2.7in,width=2.1in] {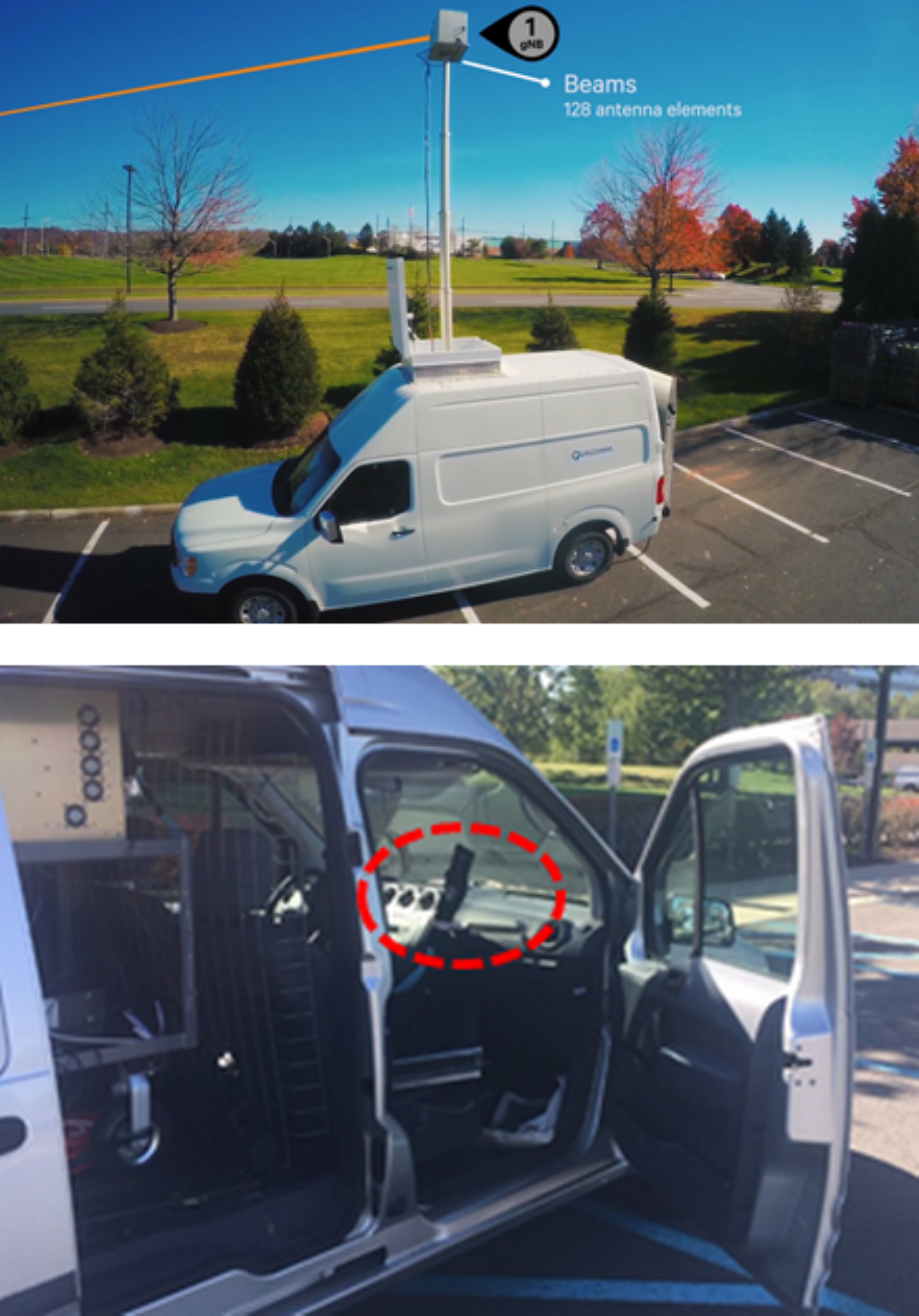}
&
\includegraphics[height=2.7in,width=3.9in] {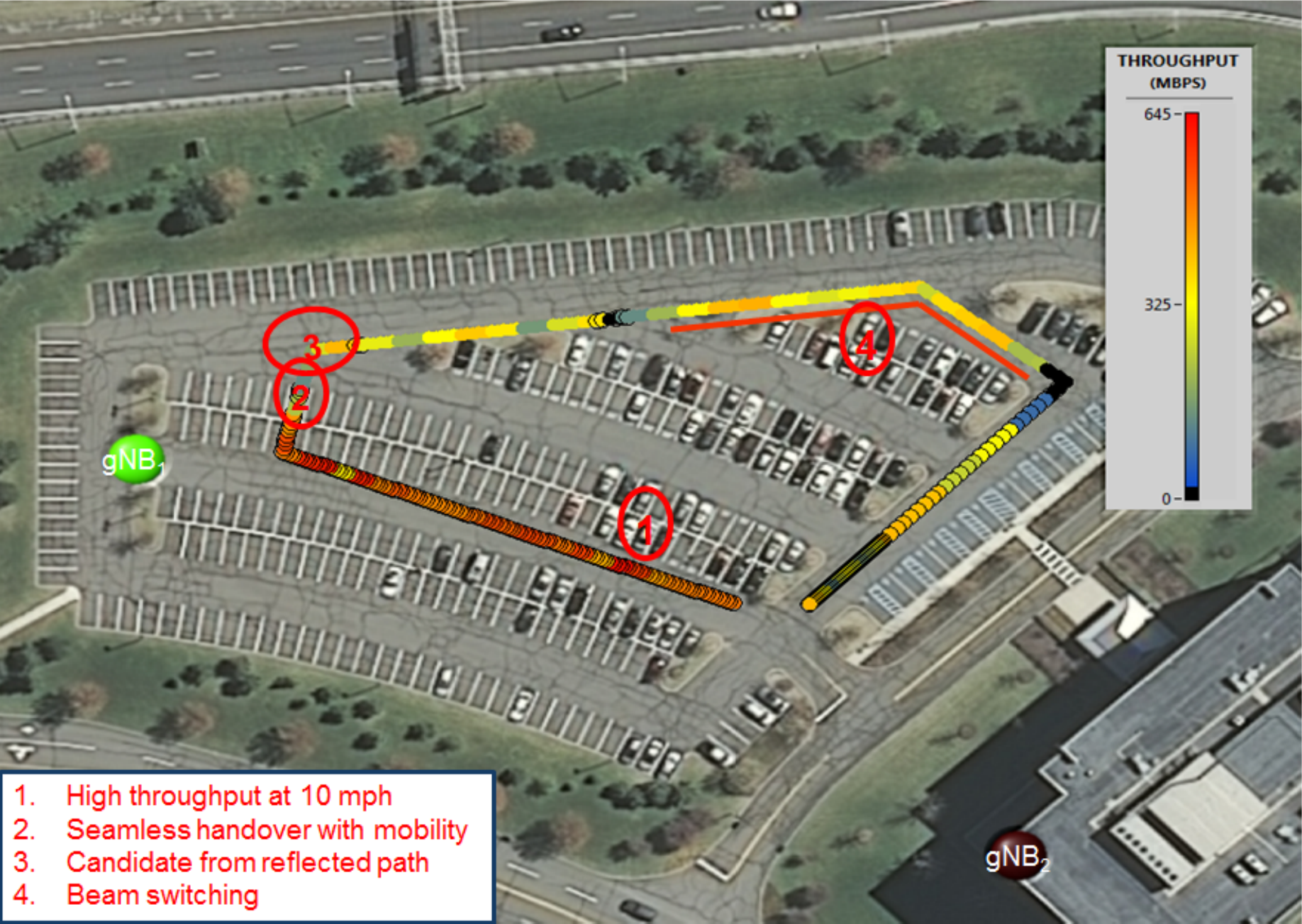}
\\
(a) & (b)
\end{tabular}
\caption{\label{fig_combined4}
(a) Elevated gNB and UE inside the testing vehicle used in outdoor testing.
(b) Aerial layout of the testing range including gNB locations, achieved rates
and important features in rates as the UE is driven over the trajectory. }
\end{center}
\end{figure*}

\section{Outdoor Mobility Studies}
\label{sec4}
An outdoor mobility testing experiment
(see Fig.~\ref{fig_combined4}(a)) is conducted in the parking spaces adjacent to the
Qualcomm building (see aerial layout in Fig.~\ref{fig_combined4}(b)). One base-station is
mounted on a mast elevated $14$ feet in a testing vehicle and located in the parking
lot (marked gNB$_1$), and another base-station is mounted in the sixth floor of the
Qualcomm building facing the window and elevated $5$ feet from the ground (marked
gNB$_2$). gNB$_1$ and gNB$_2$ have a $90^{\sf o}$ and $110^{\sf o}$ downtilt,
respectively. The UE is mounted on the dashboard of another testing vehicle and
testing is done by driving through the parking spaces at $10$-$15$ mph speeds.
Fig.~\ref{fig_combined4}(b) plots the achieved throughput (in Mbps) as a function
of the driving trajectory. From this plot, we note that a high throughput close
to $600$ Mbps is realized (Scenario $1$) when gNB$_2$ has an unobstructed LOS path
to the UE. As the UE moves over the trajectory (Scenario $2$),
seamless handover is realized between gNB$_2$ and gNB$_1$. Further, as the UE is driven
on this trajectory, the LOS path from gNB$_1$ is obstructed and communication is realized through a
reflected path first (Scenario $3$) and other paths subsequently (Scenario $4$) leading to
a drop in throughput (gradings in the heat map). This experiment illustrates the
prototype's capability to maintain
mm-Wave links robustly in outdoor scenarios with intra- and inter-base-station beam
switching and handover.

\begin{figure*}[htb!]
\begin{center}
\includegraphics[height=2.5in,width=6.6in] {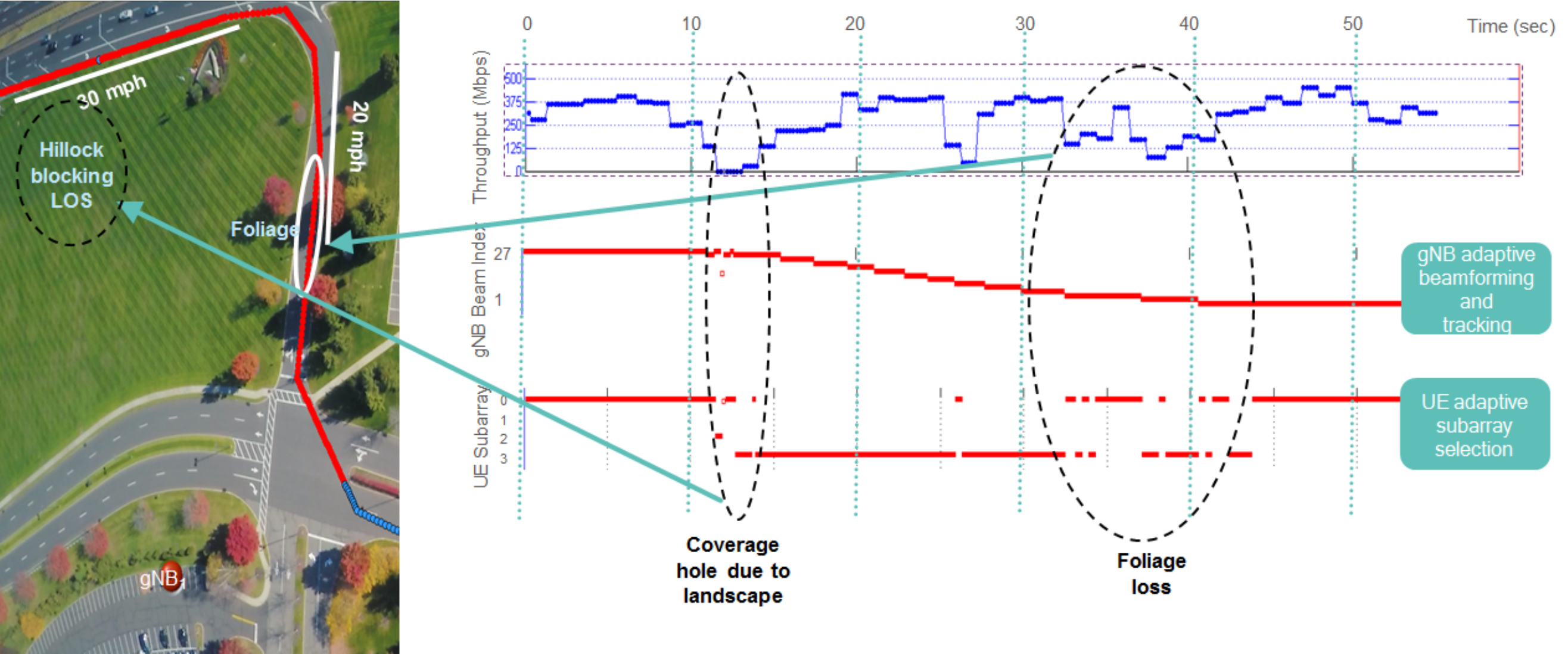}
\caption{\label{fig_combined5}
Left: Outdoor aerial layout of experiment including key geographical features.
Right: Achieved rates as a function of outdoor trajectory and time as well as
beam indices at gNB and subarray indices at UE.}
\end{center}
\end{figure*}

In a second study illustrated in Fig.~\ref{fig_combined5}, an outdoor mobility experiment
around the Qualcomm building is conducted. This environment is mostly a tree-lined
open square-type setting with some street canyon-type features. Specific points-of-interest
include parking lots and structures with bordering buildings having glass window panes,
foliage (a mix of pine and spruce trees), a large shopping mall in close vicinity
(Bridgewater Commons Mall), highways (US Rt.\ $202$), etc. For the specific experiment
reported here, a testing vehicle is driven for a period of $\approx 55$ seconds through
the exit lane of US Rt.\ $202$ at a speed of $20$-$30$ mph, onto the ramp and into a side
street enveloping the Qualcomm building (see trajectory in red in Fig.~\ref{fig_combined5}, Left
side).
In terms of notable observations from this experiment, a base-station mounted on a raised
platform at $24$ feet with a $90^{\sf o}$ downtilt (marked gNB$_1$) offers a LOS path
to the UE as it starts exiting from Rt.\ $202$ (throughput of $375$ Mbps). However, as
the UE traverses the exit lane, a small hillock-like feature blocks the LOS path leading
to a coverage hole that cannot bridged with any reasonable NLOS path from this base-station
and a significant deterioration in rate. This observation points at the necessity of
sufficient base-station density to enhance mm-Wave coverage under blockages. For example,
a base-station on the opposite side of the highway (Rt.\ $202$) could have provided
coverage to the UE over this coverage hole. As the UE crosses
this feature, a beam recovery process recovers the LOS path albeit with a different
subarray offering complementary coverage in the LOS direction leading to an improved
throughput of $375$ Mbps. Further, as the testing vehicle enters the ramp, blockage loss
due to foliage results in a throughput drop ($125$-$250$ Mbps) and two subarrays turn out
to be useful over this period. As the UE exits the ramp onto the adjoining street, the LOS
path is recovered leading to a throughput of over $375$ Mbps. The distance between
the gNB and UE varies from $50$-$100$ m over the whole experiment.

\begin{figure*}[htb!]
\begin{center}
\begin{tabular}{cc}
\includegraphics[height=2.1in,width=3.05in] {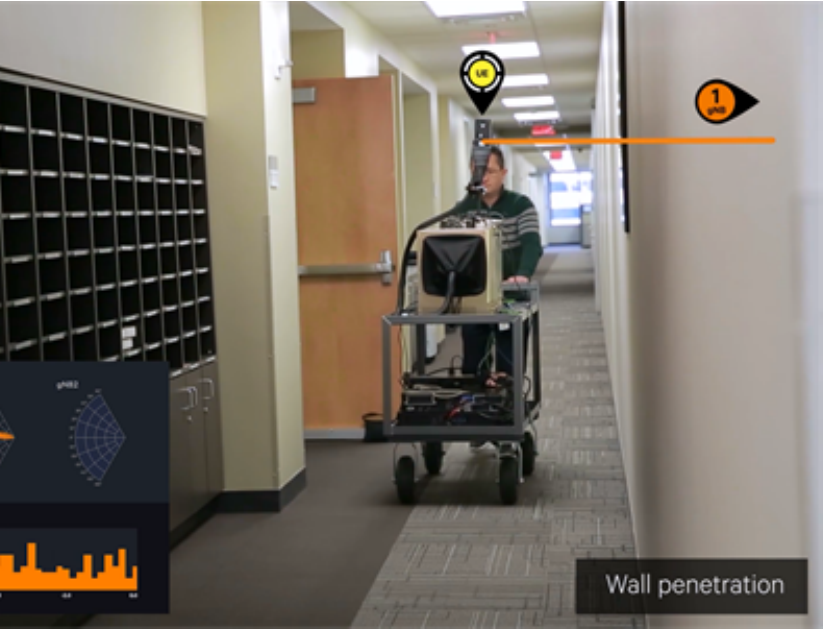}
&
\includegraphics[height=2.1in,width=3.05in] {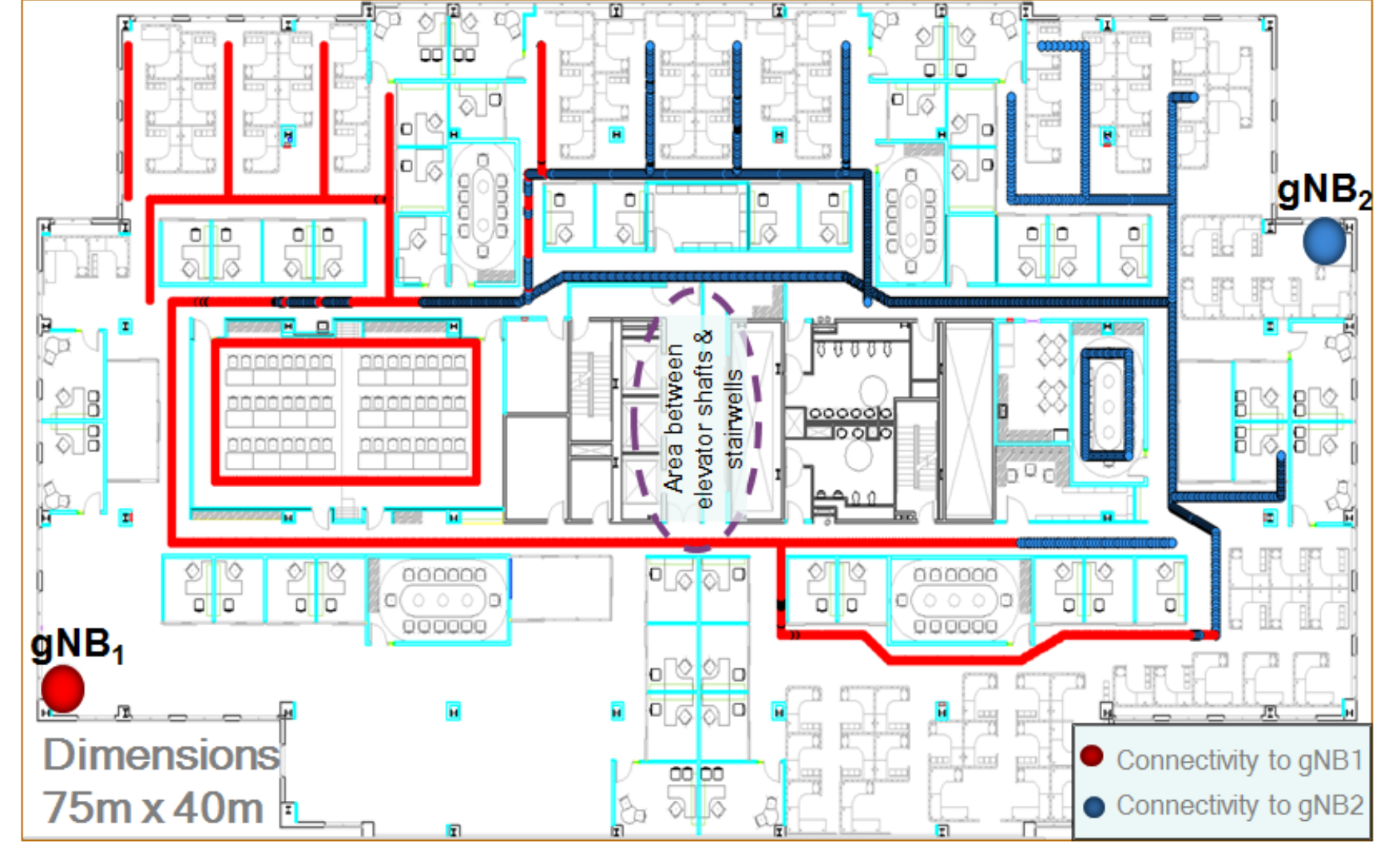}
\\
(a) & (b)\\
\multicolumn{2}{c}{
\includegraphics[height=1.0in,width=6.1in] {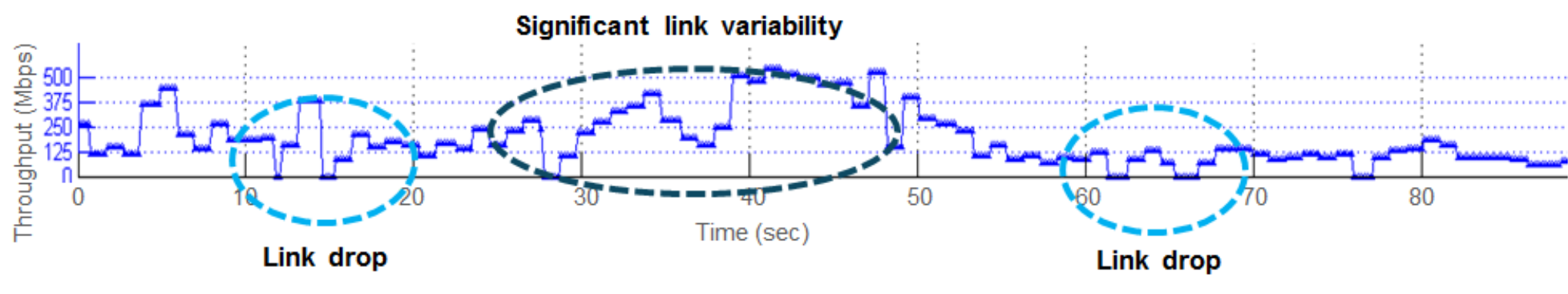} }
\\
\multicolumn{2}{c}{
(c) }
\end{tabular}
\caption{\label{fig_combined3}
(a) Typical UE testing with pedestrian mobility in an indoor scenario.
(b) Indoor layout and building plan along with two gNB locations and
coverage areas. (c) Achieved rate along with key features in the rate
trajectory over a certain indoor segment.}
\end{center}
\end{figure*}

\section{Indoor Mobility Studies}
\label{sec5}
Complementary to the above discussion,
an indoor mobility study (see Fig.~\ref{fig_combined3}(a)) in the third floor of
the Qualcomm building (see building layout in Fig.~\ref{fig_combined3}(b)) is now
described. The floor plan is mostly comprised of cubicles along the edge with
walled offices and conference rooms towards the center. Two base-stations
(marked gNB$_1$ and gNB$_2$) are placed at the far corners of the floor plan and the
UE is moved at pedestrian speeds through the layout. From our studies, these two
base-stations are sufficient to guarantee adequate coverage with at least $1$ bps/Hz
spectral efficiency as the UE is traversed through the floor plan (coverage areas with
each gNB marked in red and blue of Fig.~\ref{fig_combined3}(b), respectively).
Nevertheless, the coverage area
corresponding to each gNB does not lead to a well-defined cell boundary and is clearly
dependent on the environment, material properties, etc. This observation points to
necessity of further system coverage studies with irregular cell boundaries and
base-station density to overcome coverage holes in such scenarios. As a particular
illustration of this study, the throughput achieved with an $\approx {\hspace{-0.02in}}
90$ second trajectory
is illustrated in Fig.~\ref{fig_combined3}(c) which illustrates both link drops
due to penetration loss through obstructions (concrete, elevator area, wall,
metallic material, etc.) and link variability due to changing material properties.
Such link drops can be mitigated with enhanced beamforming, fast subarray
switching, network densification, etc.

\begin{figure*}[htb!]
\begin{center}
\begin{tabular}{cc}
\includegraphics[height=2.5in,width=3.15in] {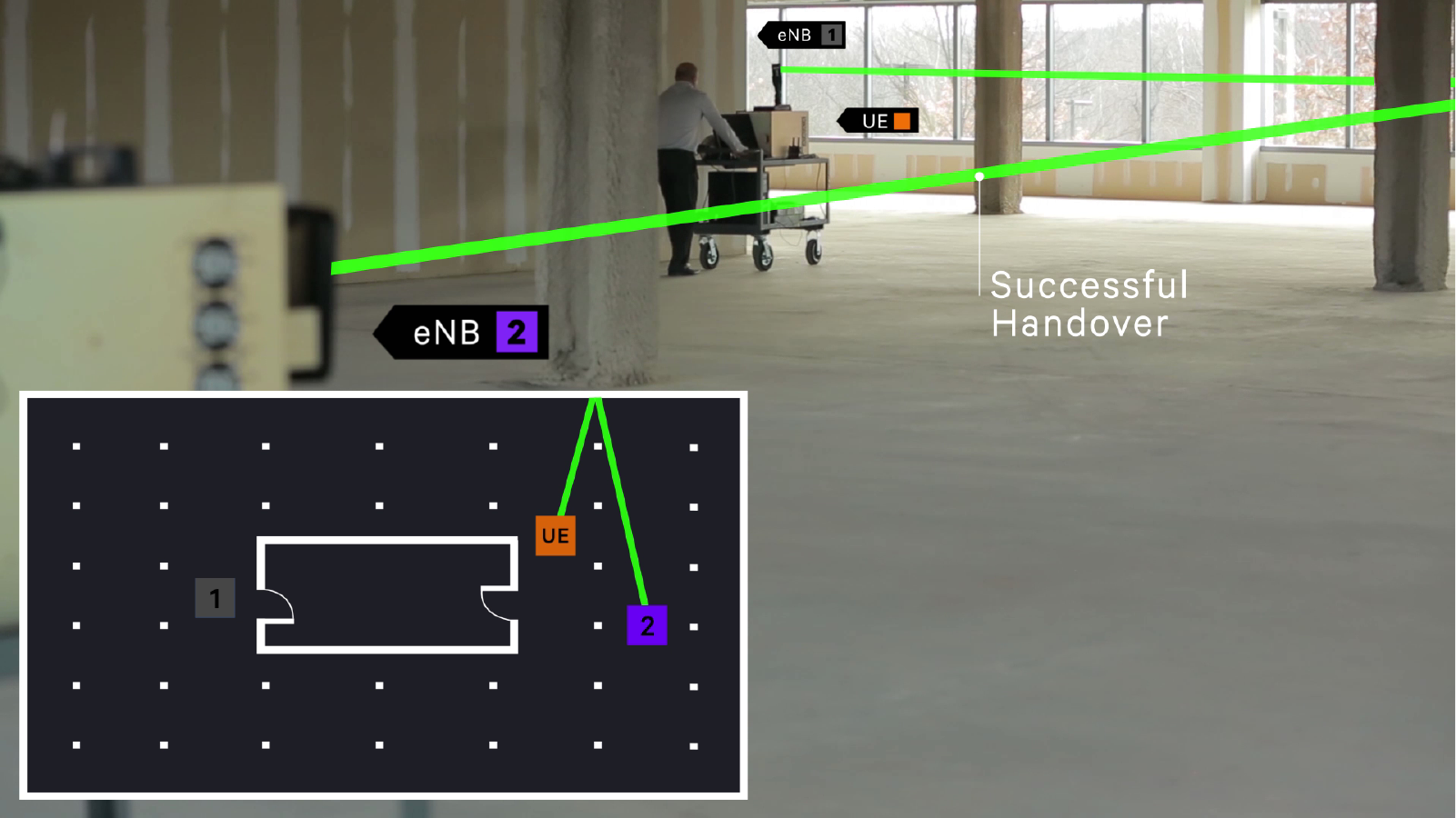}
&
\includegraphics[height=2.5in,width=3.15in] {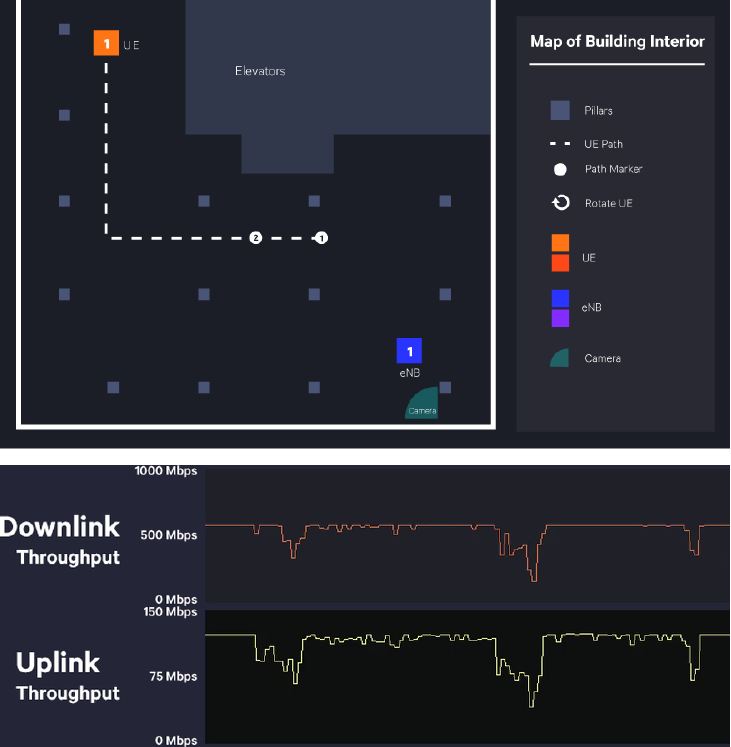}
\\
(a) & (b)
\end{tabular}
\caption{\label{fig_combined6}
(a) A successful handover from eNB$_1$ to eNB$_2$ in an indoor setup.
(b) Layout of another indoor coverage experiment with downlink/uplink rates using
the proposed beamforming solutions.}
\end{center}
\end{figure*}

More indoor mobility experiments can be seen in the video demonstration
at~\cite{qualcomm_video}. In one experiment, illustrated in Fig.~\ref{fig_combined6}(a)
and seen in~\cite{qualcomm_video}, a LOS link is initially established between the
transmitter (labeled eNB$_1$) and receiver (labeled UE) by beam scanning at both ends.
With a high ${\sf SNR}$ from the LOS link, a high rate is established on both the
downlink and uplink. As the UE is moved across the long edge of the hallway (see relative
positions of eNB$_1$, eNB$_2$ and UE in the bottom left inset of
Fig.~\ref{fig_combined6}(a)), the link connecting eNB$_1$ with the UE becomes NLOS
with increasing path loss as the distance between eNB$_1$ and UE increases. As the UE
turns the corner at the short edge, the NLOS\footnote{Note
that the LOS link between eNB$_2$ and UE is blocked by a pillar in the layout, all marked
in white squares.} link between eNB$_2$
and UE becomes better than the NLOS link between eNB$_1$ and UE and a successful
handover (illustrating robustness to blockage) happens from eNB$_1$ to eNB$_2$,
as shown in Fig.~\ref{fig_combined6}(a).

Figure~\ref{fig_combined6}(b) illustrates the layout of yet another indoor experiment where the
UE is moved from an initial position (marked ``1'' in a white circle) towards its final
destination (marked ``1'' in an orange square) via the dashed white-line trajectory. As
the UE is moved over the trajectory, the achieved downlink and uplink rates show many
disruptions as the connected path is blocked by the pillars in the layout (marked as gray
squares). For example, when the pillar blocks the LOS path, connection is established to
the dominant NLOS path (again through reflections) leading to the first
disruption in rate(s). Connection is re-established to the LOS path as the UE moves past
the pillar until the next pillar is reached where the second disruption happens. The third
disruption corresponds to the switch from the re-established LOS path to the dominant NLOS
path as the UE turns the corner. Thus, these examples illustrate the robustness of 
our proposed beamforming solutions to blockages in real deployments.

\section{Concluding Remarks}
\label{sec6}

\subsection{Summary}
This article provides a brief overview of mm-Wave channel measurements and what the implications
of these measurements are for system design. An immediate consequence of the blockage
and penetration losses inherent and specific to mm-Wave systems are the poor link margins.
These motivate the necessity to reap spatial array gains via the use of
near-optimal beamforming solutions over large antenna arrays. Prominent challenges in this
goal include the limited range and performance of mm-Wave components, as well as the robustness
of the beamforming solution to spatio-temporal channel variations and its impact on overall
system design. Further, cost considerations may allow only the use of a small number of RF
chains at either end and thus the beamforming solutions should be adaptive to changes
in the RF architecture(s). Towards this goal, directional
beamforming approaches can be used as robust, low-complexity, near-optimal solutions
that help overcome the
high propagation losses at mm-Wave frequencies. Such solutions are demonstrated with our
experimental prototype, illustrating the viability of mm-Wave systems for high data-rate
requirements. In particular, the prototype system demonstrates: i) beamforming and beam
scanning, ii) outdoor coverage and mobility, iii) resilience to blockage of paths,
iv) inter-base-station handover, v) indoor mobility, and vi) interference management
in both outdoor and indoor settings.

\subsection{Future Research Directions}
Important issues that require further study include: i) a more exhaustive study on
realistic channel modeling for mm-Wave propagation, ii) models for spatio-temporal channel
variations, iii) models for impairments such as hand/body/human blockages, phase noise,
etc., iv) advanced MIMO techniques for both single- and multi-user multi-carrier
transmissions, v) impact of mm-Wave channel properties on mm-Wave system/network design issues
such as coverage and network latency tradeoffs, mm-Wave handover, interworking with sub-$6$
GHz bands and applications, integrated access-bachkaul solutions, vi)
advanced MIMO RF architectures such as~\cite{brady_tcom,rui_zhang,lens_array1,lens_array2}
for prototype studies and real deployments, vii) RF tradeoffs in form factor UE design, etc.


{\vspace{-0.05in}}
\bibliographystyle{IEEEbib}

\end{document}